\documentclass[utf8]{frontiersSCNS} 

\usepackage{url,hyperref,lineno,microtype}
\usepackage[onehalfspacing]{setspace}

\def\keyFont{\fontsize{8}{11}\helveticabold }
\def\firstAuthorLast{Bettoni {et~al.}} 
\def\Authors{D. Bettoni\,$^{1,*}$, R. Falomo\,$^{1}$, J. K. Kotilainen\,$^{2,3}$ and K. Karhunen\,$^{3}$, }
\def\Address{$^{1}$ INAF - Osservatorio Astronomico di Padova, Italy\\
$^{2}$ Finnish Centre for Astronomy with ESO (FINCA), University of Turku, Finland\\
$^{3}$ Tuorla Observatory, Department of Physics and Astronomy, University of Turku, Finland
}
\def\corrAuthor{Daniela Bettoni}

\def\corrEmail{daniela.bettoni@oapd.inaf.it}

\begin{document}
\onecolumn
\firstpage{1}

\title[quasar Environment]{{ \bf On the role of the environments and star formation for quasar activity}} 

\author[\firstAuthorLast ]{\Authors} 
\address{} 
\correspondance{} 

\extraAuth{}

\maketitle

\begin{abstract}

\section{}
We investigate the host galaxy and environment properties of a sample of 400 low z ($<$0.5) quasars that were imaged in the SDSS Stripe82. We can detect and study the properties of the host galaxy for more than 75\% of the data sample. We discover that quasar are mainly hosted in luminous galaxies of absolute magnitude $M^* -3 < M(R) < M^*${\bf \footnote{$M^*$ and $L^*$ are the characteristic
absolute magnitude and luminosity of the Schechter luminosity function of galaxies \citep{SH}}} and that in the  quasar environments the galaxy number density is comparable to that of inactive galaxies of similar luminosities. For these quasars we undertake also a study in u,g,r,i and z SDSS bands and again we discover that the mean colours of the quasar host galaxy it is not very different with respect to the values of the sample of inactive galaxies.  For a subsample of low z sources the imaging study is complemented by spectroscopy of quasar hosts and of close companion galaxies.  This study suggests that the supply and cause of the nuclear activity depends only weakly on the local environment of quasars.  Contrary to past suggestions, for low redshift quasar there is a very modest connection between recent star formation and the nuclear activity.

\tiny
 \keyFont{ \section{Keywords:} Galaxies active, quasar} 
\end{abstract}

\section{Introduction}
Quasar phenomenon assumes that the nuclear activity can be due to the major merger of two gas-rich galaxies that feed the central engine and enable the growth of a spheroidal stellar component. However, important details on the mechanisms that triggers the gas fueling and how nuclear activity influence the successive evolution of the host galaxies remain poorly understood. The study of correlations among black hole masses, properties of the host galaxies and their environments may provide relevant clues to investigate the fundamental issues on quasar activity and its role on the evolution of galaxies. Simulations \citep[e.g.][]{Menci,Hop} suggest that minor and major merging events may have a key role for triggering and fueling the nuclear/quasar activity. These effects strictly
depend on the global properties of the galaxy environment \citep{KH,DiM}. 

The quasars, at low-redshift, follow the large-scale structure traced by galaxy clusters 
but they eschew the very centre of clusters \citep{SCC1,SCC2}. On the other hand the quasar environment, on 
small scales (projected distance $<$ 0.5 Mpc), appear overpopulated by blue disc galaxies having a significant star formation rate \citep{CL1,CL2}. At higher redshifts, quasars are in some cases  associated with richer environments \citep{HG,Djorg} but also examples of modest galaxy environments are observed.  The comparison of the environments of quasars, at Mpc scales, to those of galaxies has given contradictory results partially due to small samples and non homogeneous datasets. Early studies suggested that the galaxy environment of  quasars is more strongly clustered than that of inactive galaxies \citep[e.g.][]{SBP}, while later studies based on surveys such as the Two Degree Field (2dF) and the Sloan Digital Sky Survey (SDSS) \citep{abazajian09} found  galaxy densities around quasars and inactive galaxies to be comparable to each other \citep[e.g.][]{SBM,Wake}.  

More recent studies using the SDSS archives such as \citep{Serber} and \citep{Strand}
have taken advantage of the large datasets provided by the survey to study
quasars at z $<$ 0.4. Both studies found that quasars are on average located in higher local
over-density regions than typical L* galaxies, and that within 100 kpc from the quasar the density 
enhancement is strongest. \citet{Serber} also claimed that: "high luminosity quasars
have denser small-scale environments than those of lower luminosity". These studies are, however, limited by the deepness of the surveys that do not enable to study the galaxy population much fainter that L*. 
Meanwhile, a study by \citet{Bennert} found that many low-z quasars show signs of recent or on-going
interactions with nearby galaxies, suggesting a connection between the environment and the 
quasar activation.  In spite of these imaging studies our knowledge of the physical association of the galaxies around quasars and their dynamical properties remains uncertain. 

To overcome these issues we recently carried out a systematic study of the the host galaxy and environment properties of a large ($\sim$ 400) sample of low redshift (z $<$ 0.5) quasars based on the deep images available from the SDSS Stripe82 survey.  These co-added images are $\sim$2  magnitudes more profound with respect to the standard Sloan data and give the possibility to study both the quasar hosts and the immediate environments. We are able to resolve the host galaxy for $\sim$ 300 targets \citep{F14} and to investigate the the large scale galaxy environment of all quasars \citep{K14}. For this work we adopt the concordance cosmology with H$_0$ = 70 km s$^{-1}$ Mpc$^{-1}$, $\Omega_m$ = 0.3 and $\Omega_\Lambda$ = 0.7.

\section{The sample and analysis}
\begin{figure}[h!]
\begin{center}
\includegraphics[width=13cm]{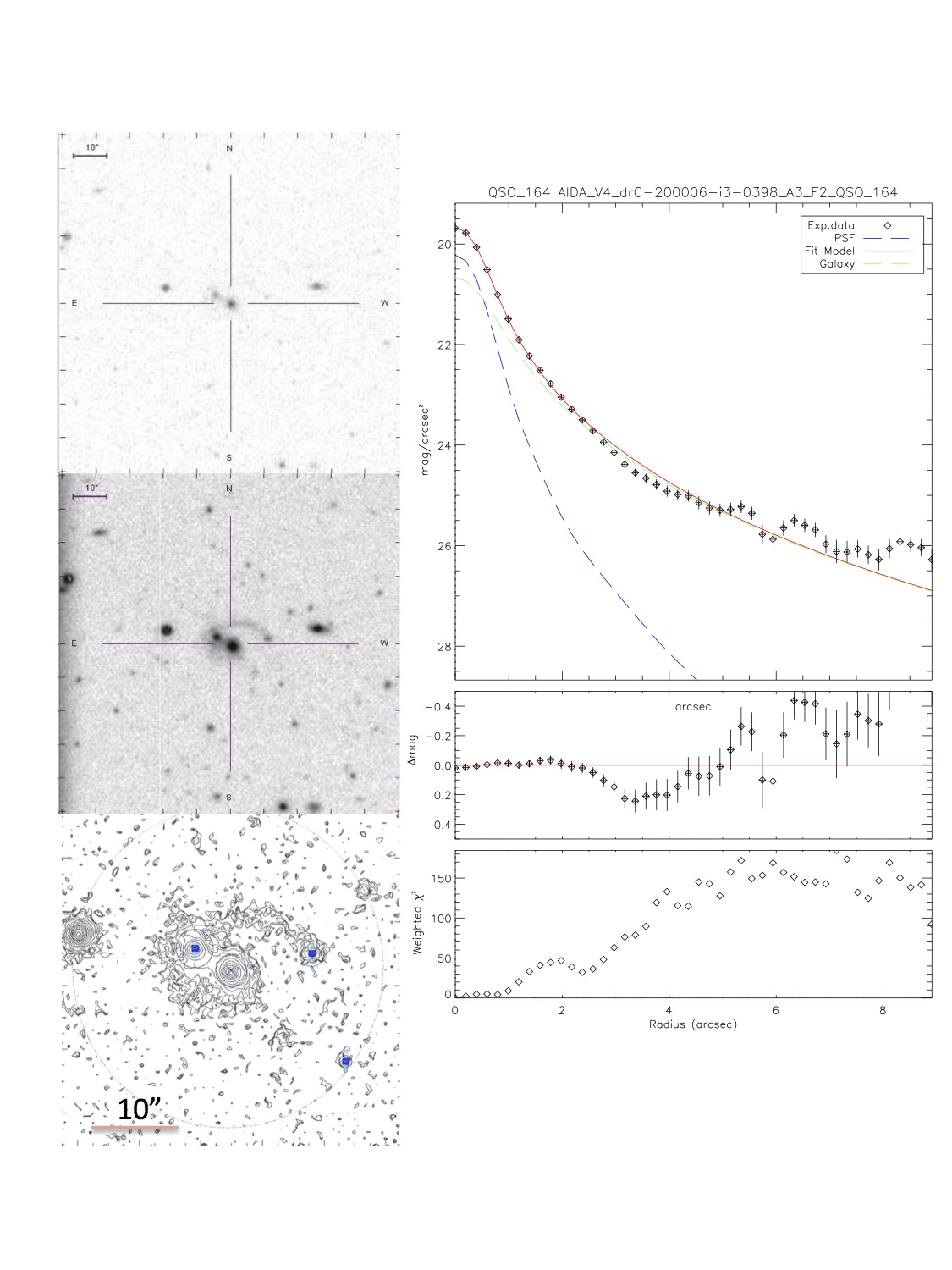}
\end{center}
\caption{Upper left panel:SDSS-DR7 i-band image, middle left panel: corresponding data from Stripe 82, bottom left panel: contour plot zoomed on the central object of the Stripe 82 i-band image (north at top, east left). Upper right panel: i-band radial profile together with the AIDA decomposition and fit. Middle right panel: distribution of residual for the best fit. Lower right panel: distribution of $\chi^2$.}\label{fig:aida}
\end{figure}

The sample of quasar we used, described in detail in \citet{F14}, is derived from the fifth release of the SDSS Quasar Catalog  \citep{schneider2010} that is based on the SDSS-DR7 data release \citep{abazajian09}.  Our analysis is done only in the region of sky covered by the stripe82 data, these images go deeper of about $\sim$2  magnitudes with respect to the usual Sloan data and make possible the study of the quasar hosts, with these images we reach $m_{i(limit)}$=24.1 \citep{Annis}. The final sample is composed by 416 quasars in the range of redshift 0.1$<z<$0.5. In this sample we are dominated by radio quiet quasars only 24 are radio loud (about 5\%). The mean redshift of the sample is $<z>$ = $0.39\pm0.08$ (median $0.41\pm0.06$ ) and the average absolute magnitude is :$<M_i>$ = $-22.68\pm0.61$ (median $-22.52\pm0.35$).
To perform the analysis of the images we used the tool AIDA (Astronomical Image Decomposition \& Analysis) \citep{Uslenghi}. The tool is designed to perform 2D model fitting of quasar images including Adaptive Optics data and with detailed modeling of the PSF and its variations. In figure \ref{fig:aida} we show an example of our analysis and of the AIDA fit.

To perform the analysis of the large scale environment we used SExtractor \citep{BA} to create our own galaxy catalogs for each image, measuring the magnitudes of the galaxies in each of the five filters separately. 
A comparison of the catalog generated with SExtractor to those of SDSS, and a visual inspection on
a number of frames to further to study the validity of our classification is fully described in \citep{K14}.  
We  found  a  good  match  between  the  SDSS and SExtractor catalogs at apparent magnitudes $m_i <$23, but  at  fainter  magnitudes  the  number  of  objects  detected by SExtractor dropped dramatically compared to those in the SDSS catalog. A visual inspection of these faint objects showed that they are mainly background noise which is either undetected by SExtractor or classified as an unknown object. We chose a conservative limit of
classification value $\leq$0.20 for our galaxies, to make sure we avoid all the stars and majority of the unknown objects.

\section{Host Galaxies}
\begin{figure}[h!]
\begin{center}
\includegraphics[width=9cm]{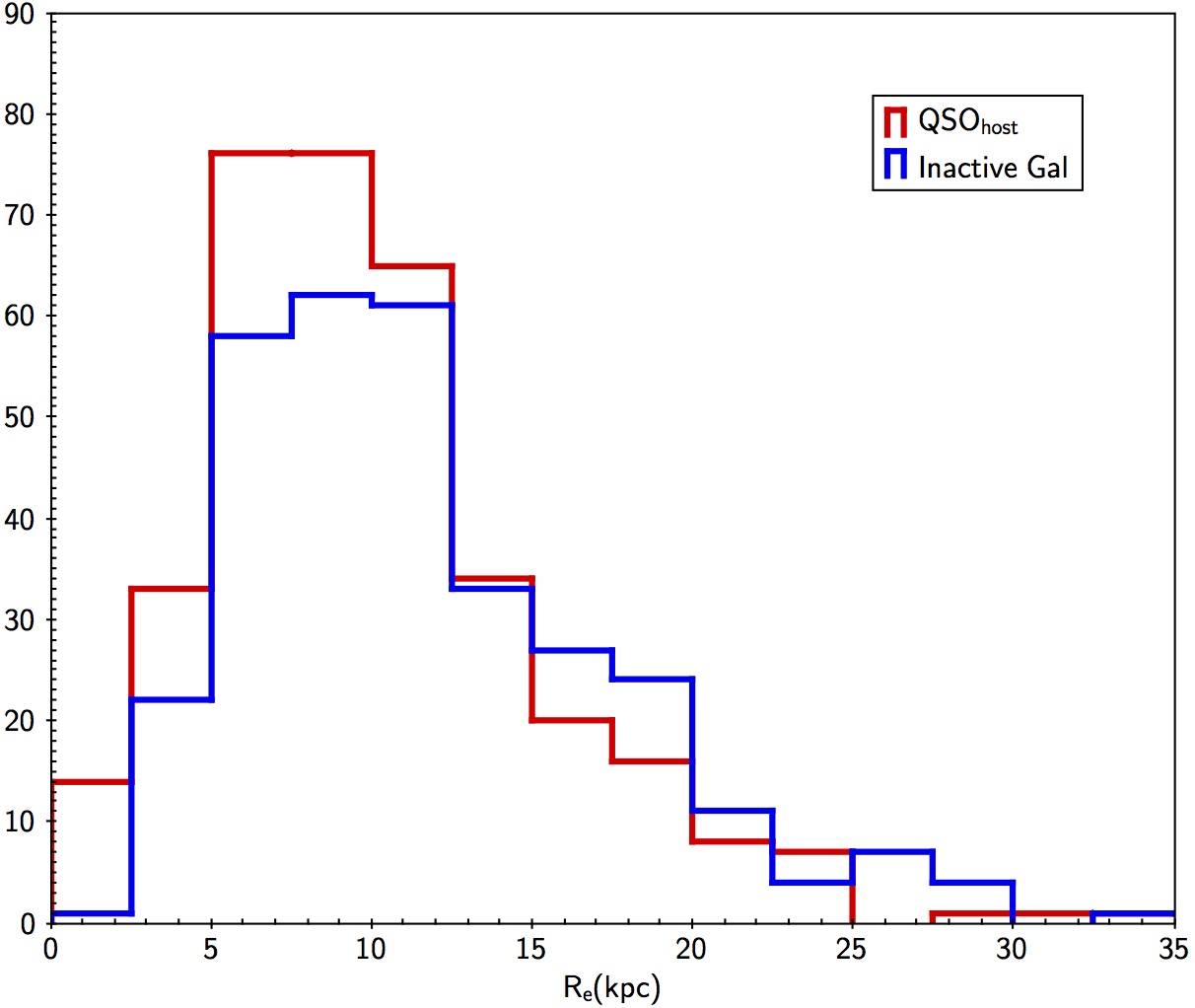}
\end{center}
\caption{Distribution of effective radius $R_e$ for quasar hosts (red line) and that of a sample of inactive galaxies (blue line) 
that have similar luminosity and redshift of quasar hosts.}\label{fig:re}
\end{figure}

For our sample we found that the morphology of the quasars host galaxy ranges from pure ellipticals to complex/composite morphologies that combine disks, spheroids, lens and halo and are dominated by luminous galaxies with absolute magnitude in the range M*-3 $<$ M(R) $<$ M*  and the average absolute magnitude of this sample is $<M_i> = -22.68 \pm 0.61$. The galaxy sizes (defined here as the half-light radius) range from very compact (few kpc) up to more extended galaxies (10-15 kpc) and, in our redshift range,  not significant trend of change of the galaxy size with z is found. However, note that the sampled redshift range is small thus possibly hiding systematic changes of size with z. It is interesting to note that the distribution of sizes of the quasars hosts is very similar to that of the sample of inactive galaxies of similar luminosity and redshift distribution as shown in figure \ref{fig:re}. The black hole mass of the quasar, estimated from the spectral properties of the nuclei, are poorly correlated with the total luminosity of the host galaxies  but  the relation between M$_{BH}$ and the luminosity of the spheroidal component is consistent with that of local inactive galaxies  \citep{F14}. We can resolve the host galaxy in $\sim$73\% of the quasar sample but in u-band only for the 48\%.

Only the ($g-i$) colour is slighter bluer (1.06 $\pm$ 0.11) than that of inactive galaxies (1.19$\pm$ 0.25) of similar luminosity and redshift \citep[see][]{B15}. The average absolute magnitude ($<M_i>_{QSOHOST}$ = -22.52 $<M_i>_{GAL}$ = -22.26) are very similar  and also ($u-g$) color of quasar hosts are similar (1.40$\pm$0.30) to that of inactive galaxies (1.54$\pm$0.62). Quasar and control sample have the same distribution of close ($<$ 50 kpc projected) companion galaxies  \citep{B15}. This result is in contrast with that obtained by \citet{M14} who find  similar (g-i) color for the quasar hosts but claim they are systematically bluer than the ensemble of normal galaxies in SDSS. Because of the different selection of active and non active galaxies of our and \citet{M14} samples it is not clear what is the cause of this discrepancy.
  
\section{Galaxy environment}
\begin{figure}[h!]
\begin{center}
\includegraphics[width=9cm]{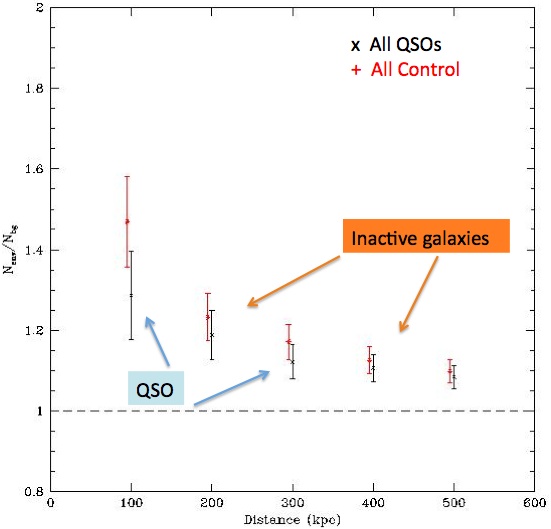}
\end{center}
\caption{Galaxy over-density  in the environment of  quasar compared with that of inactive galaxies at same redshift and luminosity.}\label{fig:env}
\end{figure}

The large scale galaxy environment ($<$1 Mpc) of the 302 resolved quasars compared with those of a sample of inactive galaxies at same redshift \citep{K14} shows that the galaxy number density is comparable to that of  inactive galaxies with similar luminosities, as seen in figure \ref{fig:env}. For distances $<$400 kpc both quasars and inactive galaxies environments shows a significant excess compared to the background galaxy density. In particular the quasars, on average, have the tendency to be associated with small group of galaxies. No statistically significant difference is found between the over densities of low z quasar around the quasars and the inactive galaxies \citep{K14}. No dependence of the over density on redshift, quasar luminosity, the galaxies at same luminosity of the host galaxy, the radio luminosity or BH mass was found. 

The lack of a notable difference between the quasars and non-active galaxies \citep{K14} environments suggests that the connection between the quasar activity and its environment is less important than believed for fueling and activate the quasar. This may also indicate that secular evolution \citep[e.g. disc instabilities, see;][] {Ciotti} may play an important role in triggering the quasar activity, and that mergers are less important than expected.

As a second step in this environment study we used all five SDSS filters to analyze the colour properties of the quasar environment (Karhunen et al. in preparation). This was done for a subsample of objects for which the SDSS-Stripe82 u-band image was deep enough to allow the detection of nearby galaxies. This implies that we have to discard $\sim$20\% of our original sample (typically u-band images with exposures $<$40 min). We find that the colours of the environments of the quasars are slightly bluer than those of inactive galaxies at all distances, with the effect being most noticeable in g-i. The colour of the environment depends on the density of the environment, with denser areas being redder. We also find that quasars at higher redshifts have redder environments.

\subsection{Close Environments}

\begin{figure}[h!]
\begin{center}
\includegraphics[width=10cm]{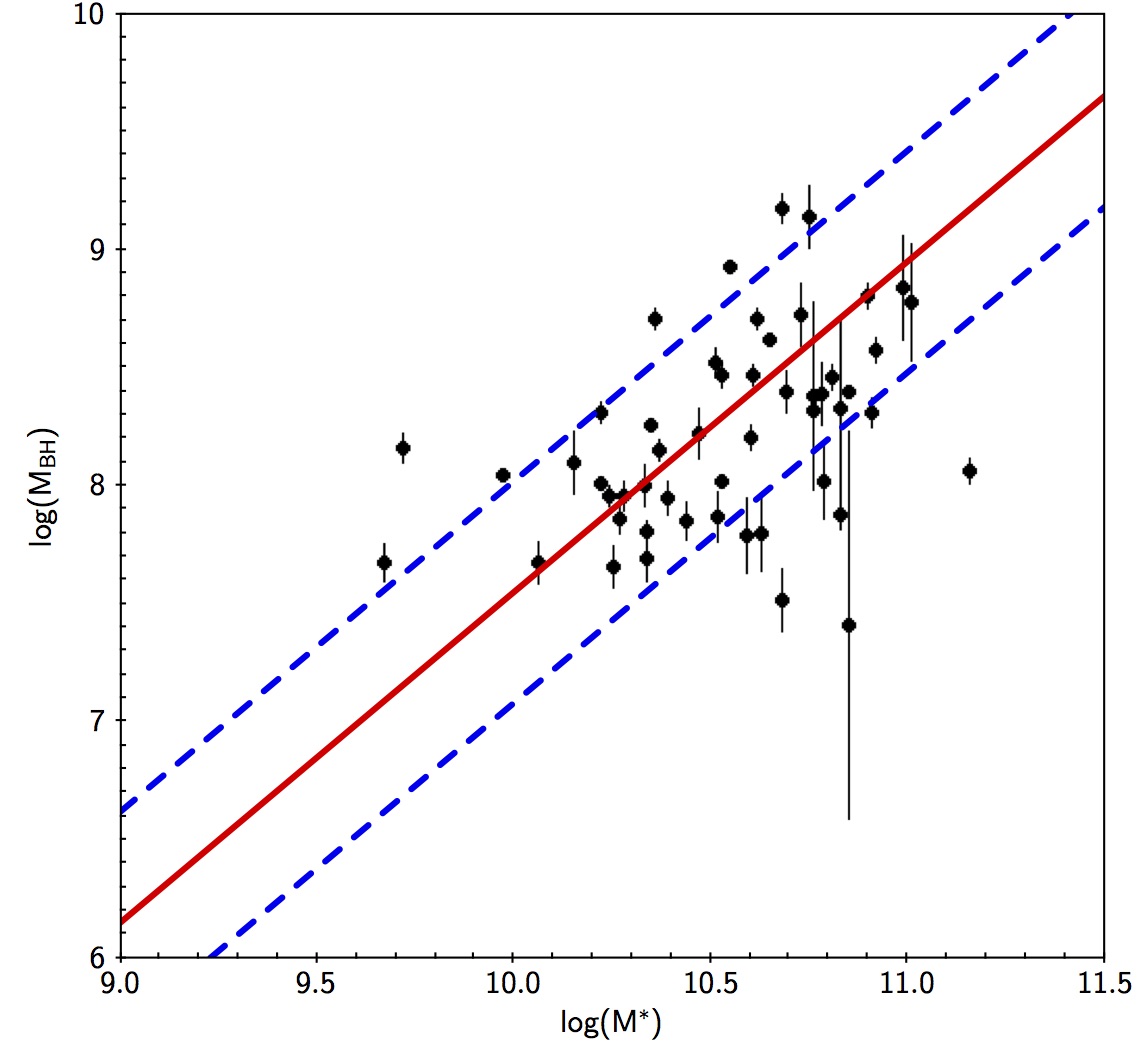}
\end{center}
\caption{Correlation between quasar virial black hole mass end the total stellar mass of their host galaxies as derived from SED fitting. {\bf Despite the large scatter the relation is similar} to that obtained by  \citet{RV}  for inactive galaxies and bulges  (red solid line). The 1$\sigma$ uncertainties are indicated by blue dashed lines.}\label{fig:BH}
\end{figure}

To better investigate, in all 5 Sloan bands, the very close environment of quasars in our sample, we selected  both from the full sample of 416 quasar and from the comparison sample only objects at z $<$ 0.3 in fact, beyond this limit, the characterization of the quasar host galaxies becomes difficult at bluer filters due to the reduced contrast between the host galaxy and the nuclear emission. The detailed description of the final sample, composed by 52 quasars, is described in \citet{B15}. To compare the colours of the host galaxy we need to evaluate the K-correction for each filter, to this purpose we used the KCORRECT package \citep{BR}. This allow us, using the luminosity+SED fitting, to derive also an estimate of the stellar mass for this subsample of quasar hosts and we used the stellar masses to derive the relation $LogM_*$-Log$M_{BH}$ shown in fig~\ref{fig:BH}. Our data are well fitted by the relation for nearby normal elliptical and S0 galaxies as derived by \citet{RV}, confirming our findings that quasars are hosted in luminous early-type galaxies.

The two samples show similar properties in particular, for our sample of resolved objects, we find an average mass of the quasar host galaxy of $<M_*> = 4.28\pm2.76\times10^{10} M_{\odot}$ and $<M_*> = 5.27\pm3.88\times10^{10} M_{\odot}$ for the comparison sample of normal galaxies. 
The overall mean colours of the quasar host galaxy are indistinguishable from those of inactive galaxies of similar luminosity and redshift. There is a suggestion that the most massive quasar hosts have bluer colours and show a higher star formation rate than those  in the sample of inactive galaxies.

\subsection{Star formation}

As a last step in our study we obtained with NOT+ALFOSC instrumentation long slit spectra of a subsample of the 52 quasar in \citet{B15} and of the close companions \citep[see details in][] {B17}. We found that in 8 out of 12 ($\sim$67\%) quasar the closest companion galaxy is associated to the quasar (same redshift). However the average level of star formation of companion galaxies that are associated with the quasar appears similar to that of the companion galaxies that are not associated with the quasar. These results suggest a modest role of the quasar emission for the SF in nearby companion galaxies. Finally for 3 targets we observed also the spectrum of the host galaxy which turned out to be typical of an old stellar population.

\section{Conclusions}

From our multi-band study of a large sample of nearby quasar we found that:

\noindent
Quasar host galaxy luminosity is mainly in the range $M^*-3 < M_R < M^*$. \\
Galaxy environments of quasar are similar to those of a sample of inactive galaxies at similar redshift and luminosity. The colours of the galaxies in the immediate environments  of the quasars are slightly bluer than those of inactive galaxies at all distances; the effect being most noticeable in ($g-i$) colour. \\
Overall the mean colours ($g-i$ = 0.82$\pm$0.26; $r-i$ = 0.26$\pm$0.16 and $u-g$ = 1.32$\pm$0.25) of the quasar host galaxy are indistinguishable from those of inactive galaxies of similar luminosity and redshift. \\
In $\sim$50\% of the quasar in the z$<0.3$ sample, we found companion galaxies at projected distance $<$50 kpc that could be associated to the quasar. \\
Optical spectroscopy of a subsample of 12 quasar shows that these associated companions exhibit [OII] $\lambda$ 3727 \AA ~emission lines suggestive of episodes of (recent) star formation possibly induced by past interactions. The rate of star formation is, however, similar to that of companions of inactive galaxies. This may indicate that the presence of the quasar do not change the star formation in the associated close companions.  Contrary to past suggestions \citep[e.g.][]{M14} the role of these associated companion galaxies for triggering and fueling the nuclear activity appears thus  marginal as similar conditions are observed in inactive galaxies of  mass similar to that of quasar hosts. Contrary to past suggestions \cite{}
A significant time delay between the phase of nuclear activity and the star formation  in the immediate environments  could smear the link between them.





\section*{Conflict of Interest Statement}

The authors declare that the research was conducted in the absence of any commercial or financial relationships that could be construed as a potential conflict of interest.

\section*{Author Contributions}

DB and RF wrote the paper and analyzed the data JK helped in the selection of the sample and the discussion of the results and KK made the NOT observations. All authors participated in the discussion of the final results.


\section*{Acknowledgments}
Funding for the SDSS and SDSS-II has been provided by the Alfred P. Sloan Foundation, the Participating Institutions, the
National Science Foundation, the U.S. Department of Energy, the National Aeronautics and Space Administration, the Japanese
Monbukagakusho, the Max Planck Society, and the Higher Education Funding Council for England. The SDSS Web Site is
http://www.sdss.org/.

The SDSS is managed by the Astrophysical Research Consortium for the Participating Institutions. The Participating
Institutions are the American Museum of Natural History, Astrophysical Institute Potsdam, University of Basel, University
of Cambridge, Case Western Reserve University, University of Chicago, Drexel University, Fermilab, the Institute for
Advanced Study, the Japan Participation Group, Johns Hopkins University, the Joint Institute for Nuclear Astrophysics, the
Kavli Institute for Particle Astrophysics and Cosmology, the Korean Scientist Group, the Chinese Academy of Sciences (
LAMOST), Los Alamos National Laboratory, the Max-Planck-Institute for Astronomy (MPIA), the Max-Planck-Institute for
Astrophysics (MPA), New Mexico State University, Ohio State University, University of Pittsburgh, University of
Portsmouth, Princeton University, the United States Naval Observatory, and the University of Washington.


\bibliographystyle{frontiersinSCNS_ENG_HUMS} 
\bibliography{ref.bib}


\section*{Figure captions}

\end{document}